\documentclass[aps,prl,twocolumn,superscriptaddress,floatfix]{revtex4-1}

\usepackage{graphicx}
\usepackage{amsmath}
\usepackage{bbold}
\usepackage[T1]{fontenc}
\usepackage{xcolor}
\usepackage{hyperref} 
\usepackage{subfigure}
\usepackage[normalem]{ulem}

\newcommand{\beq}{\begin{equation}}
\newcommand{\eeq}{\end{equation}}
\newcommand{\beqs}{\begin{equation*}}
\newcommand{\eeqs}{\end{equation*}}
\newcommand{\beqa}{\begin{eqnarray}}
\newcommand{\eeqa}{\end{eqnarray}}
\newcommand{\beqas}{\begin{eqnarray*}}
\newcommand{\eeqas}{\end{eqnarray*}}
\def\bals#1\eals{\begin{align*}#1\end{align*}}
\def\bal#1\eal{\begin{align}#1\end{align}}

\newcommand{\bcent}{\begin{center}}
\newcommand{\ecent}{\end{center}}

\newcommand{\bitem}{\begin{itemize}}
\newcommand{\eitem}{\end{itemize}}

\begin{document}

\title{Redefining the dielectric response of nanoconfined liquids:
insights from water}

\author{Jon Zubeltzu}
\affiliation{Department of Applied Physics, 
Engineering School of Gipuzkoa, Basque Country University, 
UPV-EHU, Europa Plaza 1, 20018 San Sebastian, Spain.}
  
\author{Fernando Bresme}
\affiliation{Department of Chemistry, Molecular Sciences Research 
Hub, Imperial College London, London W12 0BZ, United Kingdom 
and Thomas Young Centre for Theory and Simulation of Materials, 
Imperial College London, London SW7 2AZ, United Kingdom}

\author{Matthew Dawber}
\affiliation{Department of Physics and Astronomy, Stony Brook University, 
New York 11794-3800, USA}

\author{Marivi Fernandez-Serra}
\affiliation{Department of Physics and Astronomy, Stony Brook University, 
New York 11794-3800, USA}
                            
\author{Emilio Artacho}
\affiliation{CIC Nanogune BRTA and DIPC, Tolosa Hiribidea 76, 
             20018 San Sebastian Spain}
\affiliation{Ikerbasque, Basque Foundation for Science, 
48011 Bilbao, Spain}
\affiliation{Theory of Condensed Matter,
             Cavendish Laboratory, University of Cambridge, 
             J. J. Thomson Ave, Cambridge CB3 0HE, United Kingdom}

\date{\today}

\begin{abstract}
  Recent experiments show that the relative dielectric
constant $\epsilon$ of water confined to a film of nanometric thickness
reaches a strikingly low value of 2.1, barely above the bulk's 1.8 value
for the purely electronic response.
We argue that $\epsilon$ is not a well-defined measure for
dielectric properties at sub-nanometer scales due to the
ambiguous definition of confinement width.
  Instead we propose the 2D polarizability $\alpha_{\perp}$ as the
appropriate, well-defined response function whose magnitude can be
directly obtained from both measurements and computations. 
Once the appropriate description is used, understanding the interplay between 
electronic and ionic contributions becomes critical, contrary to what is 
widely assumed. This highlights the importance of electronic degrees of freedom 
in interpreting the dielectric response of polar fluids under nanoconfinement 
conditions, as revealed by molecular dynamics simulations. 
\end{abstract}


\maketitle

The relative dielectric constant $\epsilon_{\perp}$ across a nanometric thin film of water was recently measured to be as low as 2.1 for a width of $w \lesssim 1$ nm. This paper by Fumagalli {\it et al.}~\cite{fumagalli2018} attracted enormous attention for two main reasons: ($i$) it is a beautiful, impressive {\it tour-de-force} experiment, and ($ii$) the mentioned value is strikingly low, barely above the water bulk's purely electronic response \cite{Hill1963} $\epsilon^b_{\infty} = 1.8$. The first point is undisputed. It is the second one we want to qualify in this article.

A small value for $\epsilon_{\perp}$ is not a surprising result. Numerous simulations have already shown that values in the $\epsilon_{\perp}<10$ range are to be expected for confined water geometries \cite{Galli2013,Zubeltzu2016,DeLuca2016,Schlaich2016,Zhang2018}. The explanation of the origin of such reduced value appearing in several works \cite{Bresme2021,Michaelides2024,Deissenbeck2023} after that of Fumagalli {\it et al.} \cite{fumagalli2018}, including high-quality calculations (see e.g. Ref.~\onlinecite{Michaelides2024}), agree, in essence, with the earlier literature, all based on a combination of a more rigid structure of the water layer close to the interface \cite{Galli2013,Zubeltzu2016,DeLuca2016,Schlaich2016}, and the dead-layer effect (capacitors in series) for the thickness dependence \cite{Zhang2018}. Alternatively, in a recent work, it was proposed that the dielectric permittivity reduction of water is not connected to the structural alignment of interfacial water molecules, and instead, it emerges from anisotropic long range dipole correlations next to surfaces \cite{olivieri-JPCL-2021}. The reduction of the dielectric permittivity in other H-bonded and non H-bonded  solvents was discussed recently \cite{Motevaselian-ACS-Nano-2020}, highlighting the generality of the observed reduction of the dielectric permittivity under confinement conditions.

  The impact of interfaces on the permittivity tensor of water 
has attracted significant attention, which has stimulated
the use of the dielectric tensor as a local function
explicitly quantifying its dependence on the distance to 
the interface using statistical mechanics fluctuation 
formuli \cite{Ballenegger2005,Bonthuis2011,olivieri-JPCL-2021,
Deissenbeck2023}.
  These studies already showed that major changes in the 
dielectric response of water appear at distances $< 1$ nm from  
a surface, demonstrating the anisotropy of the permittivity tensor, 
the relevance of the surface-water interactions in determining the 
permittivity, and the importance of interfacial layers in defining 
the dielectric response of interfacial water.
Despite the intrinsic 
interest of calculating the spatial dependence of the dielectric 
permittivity profile, its interpretation in mesoscopic and molecular 
scales is not straightforward, as it requires an averaging process 
over molecular layers \cite{olivieri-JPCL-2021,Deissenbeck2023,
Michaelides2024}.
Furthermore, while most studies focused on the modification of 
molecular degrees of freedom such as dipole-dipole correlations or 
water orientation at surfaces, the impact of confinement on 
electronic degrees of freedom has escaped scrutiny.
  One aspect of particular importance, which we address in our work, 
is whether confinement influences the electronic polarisability 
of confined water relative to bulk conditions.
  This is an important question, especially considering the very low 
permittivity of confined water inferred from experiments and theory.  
The importance of confinement effects and 
of molecular degrees of freedom has been discussed more widely in the investigation of fluid flow at the nanoscale.
Advances in this area are promising, particularly in the 
development of  novel devices with enhanced properties, 
which rely to some extent on the breakdown of the macroscopic 
theories that assume a continuum framework \cite{boya-phys-today-2024}.  
  We argue below that using macroscopically defined properties 
is also problematic when considering the electrostatic response 
of fluids under nanoconfinement conditions.

  Our purpose here is not improving on the determination of $\epsilon_{\perp}$ 
but instead stating that it is an ill-defined quantity in the sub-nm regime, 
and that insisting on its quantification is misleading the discussion 
of the key variables determining the electrostatic screening of confined 
fluids. 
  As noted above, a problem emerges when assigning a dielectric 
permittivity to a molecular layer. 
  This calculation requires the definition of a layer thickness
to integrate the local perpendicular dielectric permittivity.
  The minima of the water center of mass density profiles 
normal to the surface are often used to define such thickness. 
  As discussed before \cite{fumagalli2018,Michaelides2024}, the actual 
value of the computed or measured $\epsilon_{\perp}$ is rather arbitrary, 
as the result  depends on establishing an effective film width $w$ that
cannot be uniquely defined.
  Hence, the definition of the film width becomes essentially the answer to an ill-posed question, which may have sensible answers 
but not observable ones. Think, e.g. asking what is the width of an isolated graphene sheet. In essence, this is a problem arising from carrying a 
macroscopic-theory language over to the sub-nm regime.

\begin{figure}[t!]
    \centering
    \subfigure[]{
    \hspace{2.0 mm}
        \includegraphics[width=0.4\textwidth]{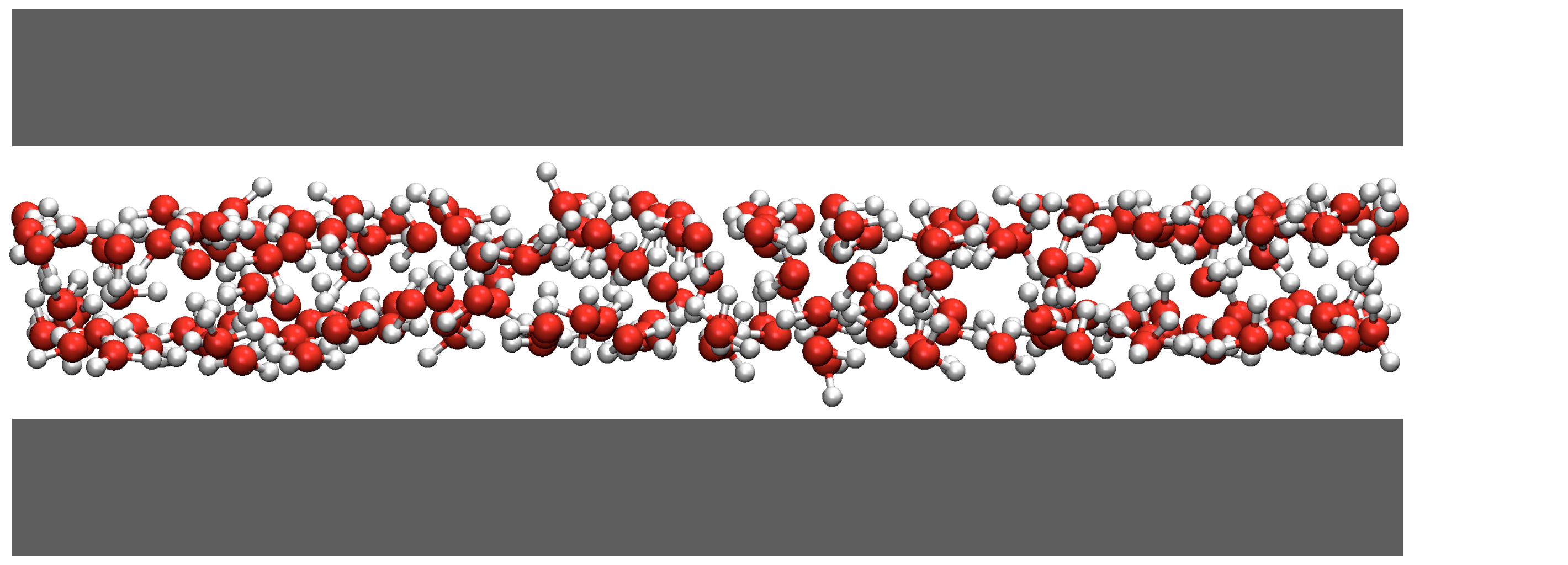}
        \label{fig:a}
    }
    \par
    \subfigure[]{
        \includegraphics[width=0.4\textwidth]{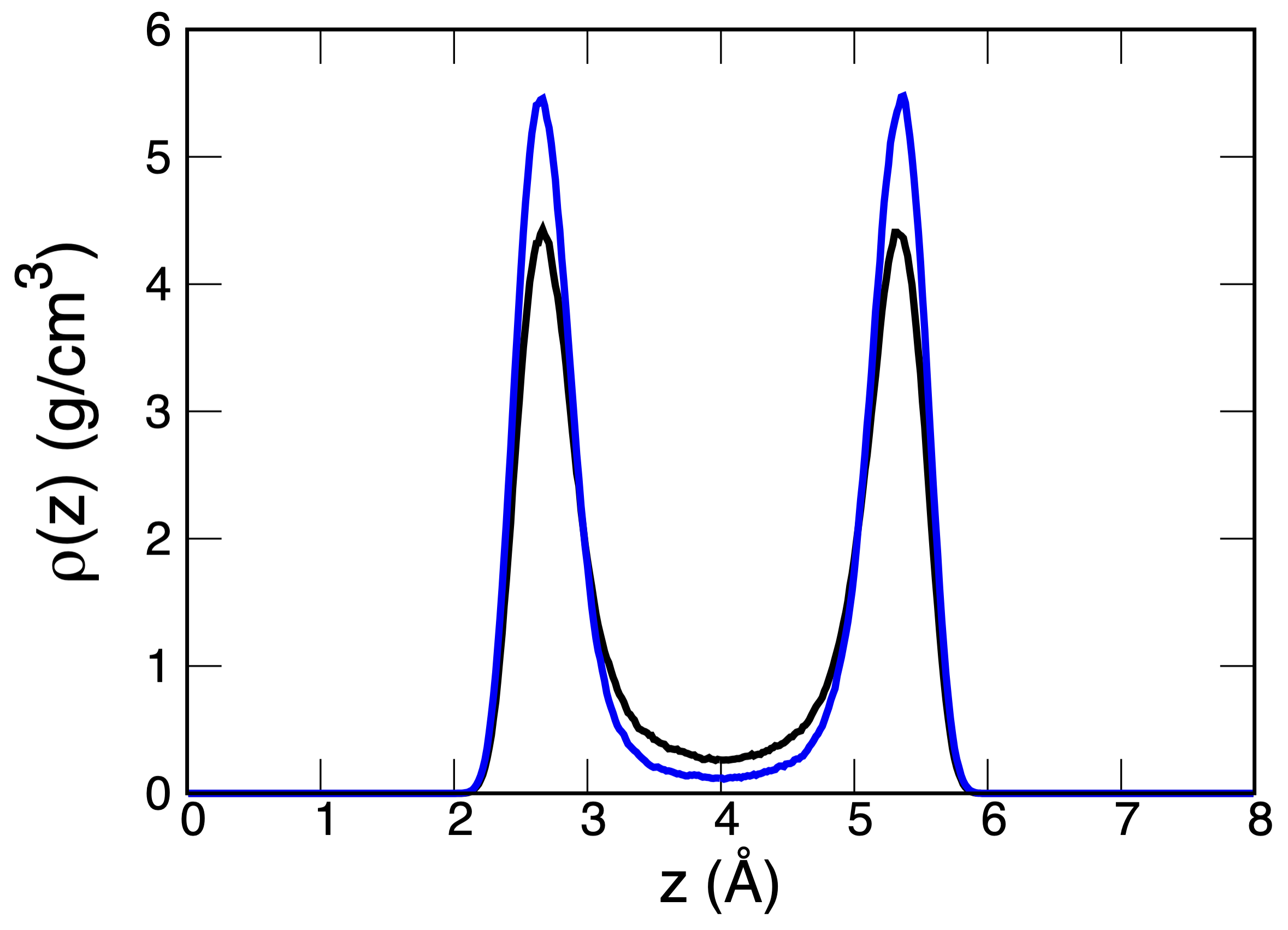}
        \hspace{10.0 mm}
        \label{fig:b}
    }
    \caption{(a) Schematic view of the simulation box, perpendicular to
    the confining plane. 
    The gray-shaded regions represent the location of the confining 
    soft potentials.
    Oxygen and hydrogen atoms in the water molecules are represented as 
    red and white spheres, respectively. (b) Oxygen density profiles along the 
    confining direction for 2D densities $\sigma=$ 0.177 \AA$^{-2}$
    (black) and 0.195 \AA$^{-2}$ (blue).}
    \label{fig:density}
\end{figure}

  The dielectric constant, relating the polarisation $P$ to the applied 
electric field, quantifies the response of a material within the 
macroscopic framework of Maxwell's equations. 
It is $P$ that is ill-defined due to the ambiguity in the width, as $P$ is given by the dipole moment per unit volume.
  Instead, the corresponding characterization for a two-dimensional 
(2D) system is given by the dipole moment per unit area, $\mathcal{P}_{2D}$. 
  We propose that for sub-nm thickness, the 2D description
is the correct one to use, the cross-over from the 3D description
being determined by the ambiguity in $\epsilon_{\perp}$ becoming 
comparable with the value itself.  
  The transverse dielectric response is then characterized by the 
2D transverse polarizability
\beq
\label{eq:def-alpha}
\alpha_{\perp} \equiv 
\frac{1}{\epsilon_0}\frac{\partial \mathcal{P}_{\mathrm{2D}}}{\partial 
\mathcal{E}^{\mathrm{ext}}_{\perp}} 
\eeq
as already proposed for 2D materials \cite{Tian2020}, instead of the 
conventional susceptibility in 3D.
  As done for the polarizability of molecules in the gas phase, 
referring to the externally applied electric field 
$\mathcal{E}^{\mathrm{ext}}_{\perp}$ is the obvious
choice here since it makes no sense to refer to the 
perpendicular field inside a 2D system.
  For solid-state thin film systems, a ``macroscopically 
averaged'' electrostatic potential can be defined as a function 
of the perpendicular component \cite{Colombo1991}, which is 
routinely used in interface and heterostructure physics.
  It has also been used in soft matter thin films (e.g. soap/water
interfaces \cite{Bresme2010}), but the difficulty remains when 
averaging the field over an ill-defined width.

  $\epsilon_0 \mathcal{E}^{\mathrm{ext}}_{\perp}$ in Eq.~\eqref{eq:def-alpha} equals to the 
electric displacement field $D_{\perp}$ to be used \cite{Sprik2016} when $P$ 
is unambiguously defined.
  The defined $\alpha_{\perp}$ has units of length and is analogous to the 
conventionally defined molecular polarisability, which has units of volume.
  It can be called the dielectric thickness of the film.
  It is important to remember key differences when referring 
to the external field instead of the internal.
  For that, the definition in Eq.~\eqref{eq:def-alpha} can be extended to 3D 
as $\epsilon_0 \alpha_{\mathrm{3D}} \equiv \partial P/ \partial 
\mathcal{E}^{\mathrm{ext}}$, using the conventional 3D polarisation $P$.
  Both polarisabilities are related by $\alpha_{\mathrm{3D}}=\alpha_{\perp}/w$
in the context of this work
\footnote{Both $\alpha_{\mathrm{2D}}$ and $\alpha_{\mathrm{3D}}$ are rank-two tensors
but we are considering one element of the diagonal of the former ($\alpha_{\perp}$),
and the latter is a number times the identity tensor for 3D bulk water.},
since $P=\mathcal{P}_{\mathrm{2D}}/w$.
  $\alpha_{\mathrm{3D}}$ relates to the conventional relative dielectric 
constant as $\alpha_{\mathrm{3D}} = 1 - \epsilon^{-1}$, which means that the
range of 1 to $\infty$ for $\epsilon$ (from no response to complete
screening of the internal field as in an ideal metal) becomes 
0 to 1 for $\alpha_{\mathrm{3D}}$.

To illustrate the main ideas conveyed here, we present two 
sets of simulations for a bilayer liquid film of water confined along the \textit{z} direction of the simulation box between two soft Lennard-Jones 9-3 separated by 8 \AA~between their respective origins (see Figure~\ref{fig:density}). The parameters of the confining potential are set to mimic the interaction of water with solid paraffin\cite{Stanley}. This same system  has been extensively studied
in previous works \cite{Stanley,Zubeltzu2016,zangi2004water,calero2020water,leoni2021nanoconfined,han2010phase}. 
The bilayer film thickness is at the lower end of the systems measured in 
Ref.~\onlinecite{fumagalli2018}. 
  For classical trajectories obtained using the empirical force field 
TIP4P/2005~\cite{TIP4P} at $T = 300$ K under various values of an applied 
external field, we calculate $\alpha_{\perp}$ from the $\mathcal{P}_{2D}$ 
obtained from ($i$) the point charges of the TIP4P/2005 model and $(ii)$ from 
the perpendicular dipole obtained with density-functional theory (DFT) 
on a sample of snapshots of the classical trajectory, under the same external electrostatic field values. The TIP4P/2005 trajectories are generated within the NVT ensemble using the Nos\'e-Hoover thermostat involving 10 ns of equilibration time followed by 10 ns of production time. The simulation box, with dimensions 34.39 \AA 
$\times$34.39 \AA 
$\times$23.00 \AA, contains between 210 and 236 water molecules. This corresponds to a 2D molecular density range of 0.177 \AA$^{-2}$ - 0.200 \AA$^{-2}$, where water is found to be in the liquid phase \footnote{The 2D molecular density range of 0.177 \AA$^{-2}$ - 0.200 \AA$^{-2}$ corresponds to a 3D mass density range of 1.04 g/cm$^3$ - 1.17 g/cm$^3$ in Ref.\cite{Zubeltzu2016}}. The periodic image interactions along the confining direction are effectively disabled by using the 
\textit{slab} command in LAMMPS \cite{LAMMPS}, with a {\tt volfactor} of 3.0.
  The DFT calculations were done with the {\sc Siesta} 
method \cite{Soler2002} using the PBE density functional \cite{PBE2}. 
  We use the TZP basis set for water described 
in \cite{corsetti2013b}. 
  To assess the accuracy of our simulation setup we determine the polarisability of a water monomer. 
We obtain a polarisability of 1.6 \AA$^3$ in Gaussian units (see Supplemental Material for calculation details), which is very close to the previously reported value of 1.59 \AA$^3$ calculated using the same functional \cite{molecule_pol}. This result is also in reasonable agreement with the experimentally reported value of 1.45 \AA$^3$ \cite{Handbook}. 
  The remaining technical details for both kinds of calculations, 
classical and DFT, as well as a detailed description of the structural 
and other properties of the film as obtained in the simulations can 
be found in Ref.~\onlinecite{Zubeltzu2016}.


\begin{figure}[t] 
\includegraphics[width=0.48\textwidth]{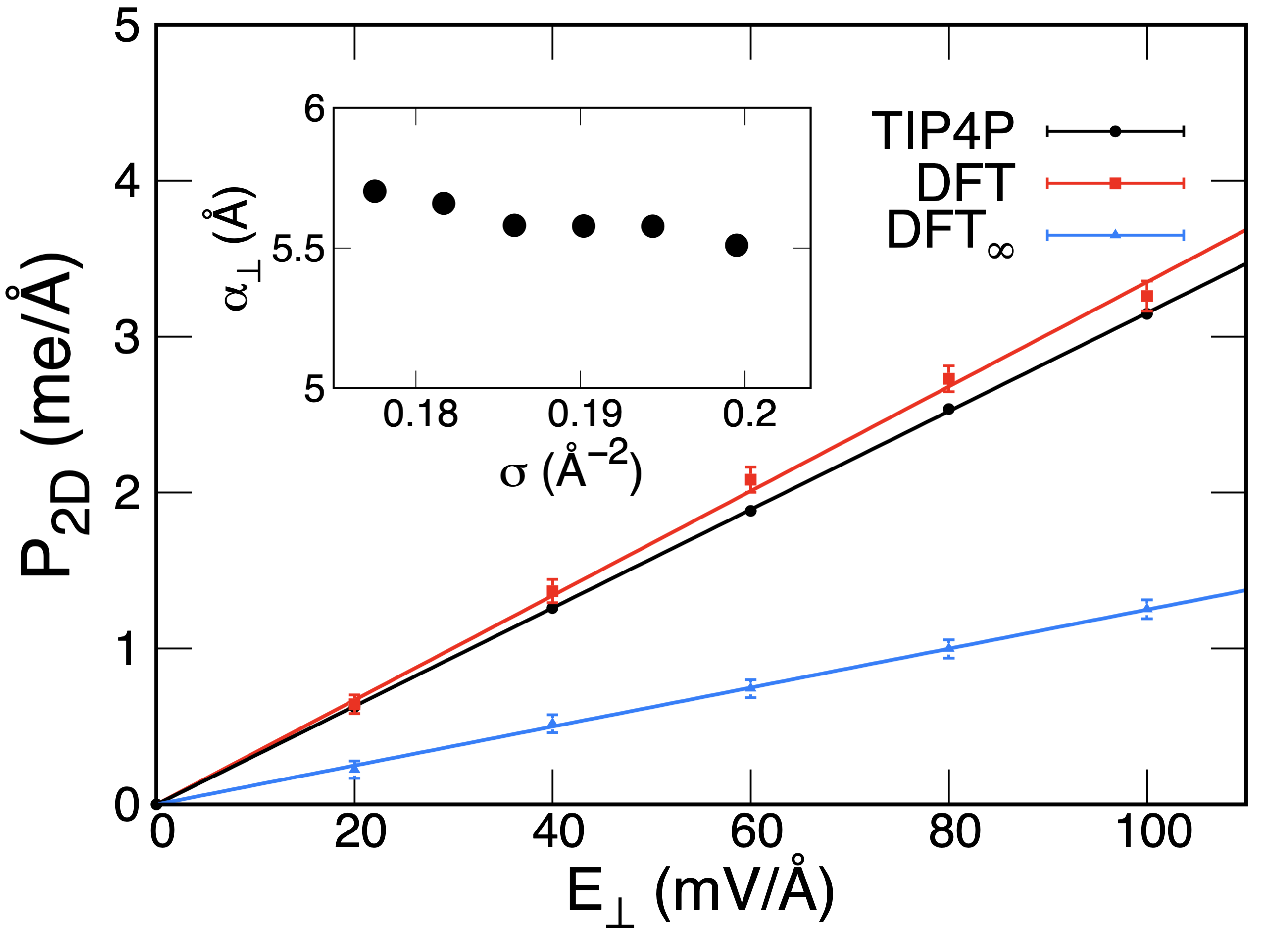}
\caption{2D polarisation $\mathcal{P}_{\mathrm{2D}}$ vs external traverse 
electric field $\mathcal{E}^{\mathrm{ext}}_{\perp}$ for TIP4P-2005 water 
classical trajectories (black) and for DFT calculations using the 
classical TIP4P-2005 under the same external field (red). 
  In blue, we show the pure electronic response obtained from 
DFT as explained in the text. Inset: 2D density dependence of $\alpha_{\perp}$ 
obtained with the TIP4P-2005 classical simulations.}
\label{fig:P-E}
\end{figure} 


\begin{table}[b] 
\centering
\caption{\label{tab:alpha}
  Two-dimensional normal polarizability of the water film in \AA. 
  The experimental value $\alpha_{\perp}^{\mathrm{exp}}$
is obtained from the values of $\epsilon_{\perp}$ and $w$ 
reported in Ref.~\onlinecite{fumagalli2018}. 
  TIP4P/2005 values ($\alpha_{\perp}^{\mathrm{TIP}}$) were obtained 
from TIP4P/2005 trajectories under various values of the normal 
electric field. 
  DFT results correspond to calculations under 
the same field values on a sample of snapshots of the 
corresponding TIP4P/2005 classical trajectories
($\alpha_{\perp}^{\mathrm{DFT}}$),
or snapshots of classical trajectories for TIP4P/2005 
under zero field conditions
($\alpha_{\perp\infty}^{\mathrm{DFT}}$). 
$\left (\alpha_{\perp\infty} / \alpha_{\perp} 
\right )^{\mathrm{exp}}_\mathrm{bulk}$ corresponds to the
ratio of polarisabilities as obtained from $\epsilon=78$ 
and $\epsilon_{\infty}=1.8$ for bulk water.}

\begin{ruledtabular}
\begin{tabular}{cccc|cc}
$\alpha_{\perp}^{\mathrm{exp}}$ & $\alpha_{\perp}^{\mathrm{TIP4P}}$ & 
$\alpha_{\perp}^{\mathrm{DFT}}$ & $\alpha_{\perp\infty}^{\mathrm{DFT}}$ &
$\left (\alpha_{\perp\infty}/\alpha_{\perp}
\right )^{\mathrm{DFT}}_\mathrm{film}$ & 
$\left (\alpha_{\perp\infty} / \alpha_{\perp} 
\right )^{\mathrm{exp}}_\mathrm{bulk}$ \\
\hline
$4.5-7.9$ & 5.7 & 6.1 & 2.3 & 0.37 & 0.45 
\end{tabular} 
\end{ruledtabular}
\end{table}

  Figure~\ref{fig:P-E} shows $\mathcal{P}_{2D}$ versus 
$\mathcal{E}_{\perp}^{\mathrm{ext}}$ 
for a 2D molecular density of $\sigma$ = 0.177 
\AA$^{-2}$, for both TIP4P-2005 and DFT calculations. Regarding the orientation of the molecular dipole moment predicted by the TIP4P/2005 model along the confining direction, two preferential orientations are observed in each water layer (see Supplemental Material). This agrees with the structure reported in previous calculations using first-principles accuracy neural network potentials \cite{Michaelides2024}.
  The 2D polarisation for the DFT calculations is slightly 
larger, consistent with the deformation of the electronic cloud. 
  The resulting values of $\alpha_{\perp}$ are shown in 
Table~\ref{tab:alpha}, and compared with the results of 
experiment \cite{fumagalli2018}.
  For the latter, there is an ambiguity since the actual 
values of capacitance, $\mathcal{P}_{2D}$, or 
$\alpha_{\perp}$ were not reported, but 
$\epsilon_{\perp}=2.1$ was given for a range 
of water film thickness between 8.5 and 15 \AA.

  The $\alpha_{\perp}$ interval appearing in the Table corresponds 
to that range of film thicknesses.
  Both theoretical values of $\alpha_{\perp}$ (TIP4P/2005, DFT) fall within
the range inferred from experiments ($\alpha_{\perp}^{\mathrm{exp}}$).
  The inset of Figure~\ref{fig:P-E} also shows the dependence of 
$\alpha_{\perp}$ on the 2D density of water molecules 
obtained from the analysis of TIP4P/2005 classical trajectories. 
  These results address the experimental difficulty in ascertaining 
the experimental pressure conditions \cite{fumagalli2018}.
  Our results  (see Fig.~\ref{fig:P-E}) show that $\alpha_{\perp}$ 
features a very small dependence on the 2D density. 

\begin{figure}[t] 
\includegraphics[width=0.47\textwidth]{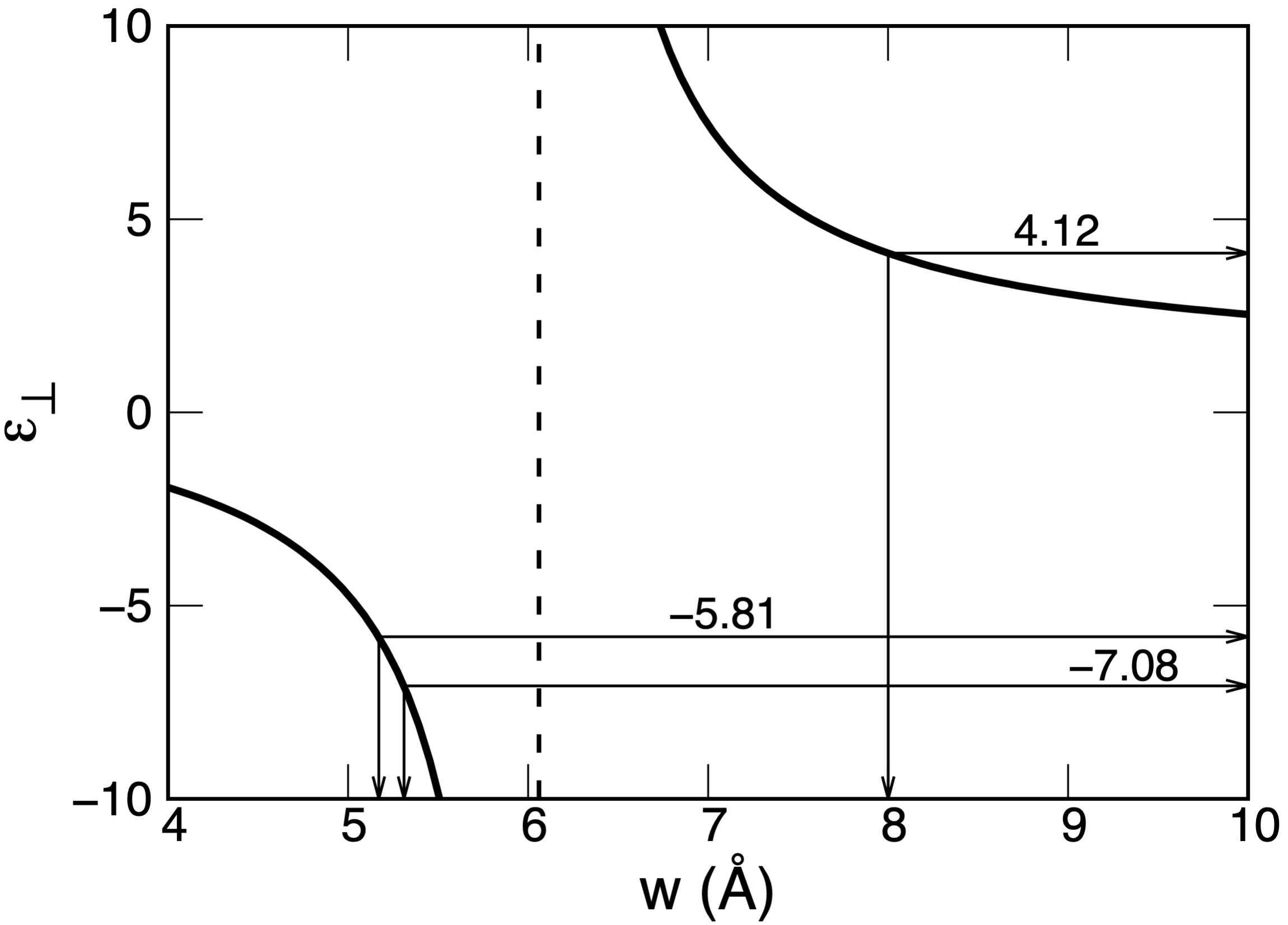}
\caption{Dielectric constant across a water thin film as a function of 
the defined width $w$ of the film for a given value of the 
2D polarisability, $\alpha_{\perp} = 6.06$ \AA, 
as obtained with DFT simulations described in the text.
  Values of $\epsilon_{\perp}$ are highlighted for three 
values of $w$ for the same film: 
  ($i$) the distance between the origins of the confining potentials $w$ = 8 \AA,
  ($ii$) $w = 5.17$ \AA, the accessible perpendicular space as 
defined by Kumar {\it et al.} \cite{Stanley}, for this bilayer film, and 
  ($iii$) $w=5.31$ \AA, obtained from normalising the 2D density of the film
with the 3D density of bulk water. 
  Negative values for $\epsilon_{\perp}$ indicate overscreening, a 
larger polarisation than what is needed to screen the field completely.}
\label{fig:epsilon-w}
\end{figure}

In addition to directly calculable, the proposed 2D 
polarisability $\alpha_{\perp}$ can be
directly obtained from experiment, in particular from
capacitance measurements using $l-\alpha_{\perp}=\epsilon_0/C$,
where $C$ is the capacitance per unit area, and $l$ is the
effective distance between the capacitor plates. 
This distance is in principle ill-defined at the nanoscale, as the
effective width of a plate capacitor depends on the charge distribution around its surface planes\cite{Marivi2}.
Differential capacitance measurements can be used to avoid the dependence 
on $l$, as done in Ref.~\onlinecite{fumagalli2018}.
In particular, by measuring the
capacitance per unit area with and without dielectric (water), $C$ and $C_0$, 
respectively,
\begin{equation}
\label{alpha-capcitance}
\alpha_{\perp}=\epsilon_0(C_0^{-1}-C^{-1}).
\end{equation}
If a change in the inter-plate distance $\Delta l$ is measured when the capacitor is filled with water (as in Ref.~\onlinecite{fumagalli2018}), it simply needs to be added to Eq. 2: $\alpha_{\perp}=\Delta l+\epsilon_0(C_0^{-1}-C^{-1})$. Note that $\Delta l$ is independent of the choice of the reference position of each plate, as it directly quantifies the displacement that occurs between the plates before and after the filling with water.
Additionally, the measurements of $C$ and $C_0$ are enough to define an ``effective" dielectric constant, $\epsilon^{eff}_{\perp}=\frac{C}{C_0}$, without the need to estimate the effective distance $l$.
$\epsilon^{eff}_{\perp}$ represents the overall dielectric behavior of the plate capacitor's interior\cite{becker2023multiscale,loche2020universal}, and it is not the dielectric constant of the water film.
It can be used to partition the system into layers and evaluate the total capacitance as a model of capacitors in series.
But such partitions, particularly the surface ones are still ill-defined and arbitrary.
We illustrate here the extreme sensitivity of $\epsilon_{\perp}$
if insisting on defining a water film width $w$, in spite of the acknowledged
arbitrariness of such a choice \cite{Michaelides2024}. 
  The macroscopic formalism is thereby recovered, with $\alpha_{\perp}$ 
relating to $\epsilon_{\perp}$ through \cite{Sprik2016}
\beq
\label{eq:eps}
\epsilon_{\perp} = \left ( 1 - \frac{\alpha_{\perp}}{w} \right )^{-1} \; .
\eeq
  $\epsilon_{\perp}$ diverges when $w=\alpha_{\perp}$, which nicely 
points to the physical meaning of the length $\alpha_{\perp}$: 
the width of an ideal metal film that would respond with the 
same polarisability as the one being measured. 
  The divergence is also responsible for the high 
sensitivity of the value of $\epsilon_{\perp}$ to the chosen $w$. 
  Fig.~\ref{fig:epsilon-w} illustrates that variability for the DFT
simulations described above, showing that a given value of 
$\alpha_{\perp}$ offers dramatically different values of 
$\epsilon_{\perp}$ for different sensible choices of $w$. 


  The 2D polarisability $\alpha_{\perp}$ as defined here is an
intensive property in the plane (per unit area), but extensive 
out of plane, and it depends on the amount of substance per unit area 
and the 2D density $\sigma$. 
  When it is normalized by the experimental density $\rho$ of bulk water 
at ambient conditions, it gives another effective thickness, 
$w = \sigma/\rho$, 
which converges to the film thickness in the macroscopic limit. 

  In our simulations, we obtain $w = 5.31$ \AA\ for the molecular 2D density
$\sigma$ = 0.177 \AA$^{-2}$ and $T=300$ K. 
  We include this thickness in the analysis shown in Fig.~\ref{fig:epsilon-w}.
  
  Finally, we address the striking experimental result that the reported 
dielectric response of the film $\epsilon_{\perp}=2.1$ is only marginally 
larger than the electronic dielectric constant of bulk water 
$\epsilon^b_{\infty}=1.8$.  

  We calculate the electronic 2D polarizability 
$\alpha_{\perp\infty}^{\mathrm{DFT}}$ of the film by running DFT 
calculations under different values of $\mathcal{E}_{\perp}^{\mathrm{ext}}$ 
on a sample of snapshots from the zero-field TIP4P/2005 trajectory. 
  The results are shown in Fig.~\ref{fig:P-E} giving the value of
$\alpha_{\perp\infty}^{\mathrm{DFT}}=2.3$ \AA\ (Table~\ref{tab:alpha}), 
which corresponds to a ratio of $\alpha_{\perp\infty}^{\mathrm{DFT}}/
\alpha_{\perp}^{\mathrm{DFT}}$ of 0.37 for the film. This ratio represents the percentage of electronic
to total out of plane polarisability in the film.
It can also be estimated for bulk water from experiments as:

\begin{equation}
    \left (\alpha^{\mathrm{3D}}_{\infty} / 
\alpha^{\mathrm{3D}} \right )^{\mathrm{exp}}_\mathrm{bulk} = 
(1-\epsilon^{-1}_{\infty}) /(1-\epsilon^{-1}).
\end{equation}

Using $\epsilon^b_{\infty}=1.8$ and $\epsilon^b=78$ the ratio is 0.45, a 40$\%$ increase with respect to the thin film.
  If insisting on keeping the 3D language to discuss the
film properties, our result indicates that if $\epsilon_{\perp}=2.1$ as 
reported in previous experiments, then $\epsilon_{\perp\infty}=1.24$  instead of the bulk accepted value 1.8.
  However, that comparison relies on the experimentally reported value for 
$\epsilon_{\perp}=2.1$, which was obtained  assuming that the film 
thickness is well defined.
  A more appropriate description of the reduction in electronic response can be established by considering the (electronic) 
molecular polarisability.
  For the film it is given by normalizing 
$\alpha_{\perp\infty}^{\mathrm{DFT}}$ with the 2D molecular density
$\sigma$ = 0.177 \AA$^{-2}$. We obtain a value of \(1.01 \, \text{\AA}^3\) in Gaussian units, which is significantly lower than the molecular polarisability of water in the bulk phase: 30\% smaller than the 1.45 \AA$^3$ inferred from experiments \cite{nir1973polarizability} and 42\% smaller than the 1.75 \AA$^3$ obtained from DFT calculations using the same exchange-correlation functional as in this work \cite{molecule_pol}.

The indications obtained in this work point to a reduction of the electronic response of the film. While this provides a plausible explanation for the striking experimental result reported, further investigation would be desirable to confirm and quantify this effect more precisely. In particular, \textit{ab initio} molecular dynamics simulations could offer valuable insight. In this sense, our observations agree with the comment of Ref.~\cite{olivieri-JPCL-2021} that the reduction in $\epsilon_{\perp}$ is not necessarily connected to interfacial water alignment. Our results suggest instead that it may be related to changes in electronic polarizability, a point that has not been considered or addressed in previous works.

That reduction becomes the intriguing result now. 
  The directional dependence of the electronic polarizability 
tensor of the water molecule \cite{molecule_pol} is too small 
to support the hypothesis of the electronic response reduction
being due to prevalent molecular orientations in the film.

  Nonetheless, the thickness dependence observed for $\epsilon_{\perp}$ 
remains nicely explained by the capacitors-in-series, dead-layer 
picture. 
  The point that it is $\alpha_{\perp}$ what is suitable for 
observation, both experimental and computational, does not 
invalidate a good phenomenological theory for analysis but 
care should be exercised when attempting to interpret the 
experimental behavior in terms of ill-defined experimental observables.


\begin{acknowledgments}
  We acknowledge the Basque Government 
for financial support through grants IT1254-19, IT1584-22 and  
SGi/IZO-SGIker UPV/EHU for computational resources.
  Funding from the Spanish MCIN/AEI/10.13039/501100011033 
is also acknowledged, through grants PID2019-107338RB-C61 
and PID2022-139776NB-C65, as well as a Mar\'{\i}a de Maeztu 
award to Nanogune, Grant CEX2020-001038-M, and the United Kingdom's EPSRC Grant no. EP/V062654/1. MVFS was funded by the U.S. Department of Energy, Office of Science, Basic Energy Sciences, under Award No. DE-SC0019394, as part of the CCS Program. FB thanks the ICL RCS High-Performance Computing
facility and the UK Materials and Molecular Modelling Hub, partially funded by the EPSRC
(Grant Nos. EP/P020194/1 and EP/T022213/1). We thank Dr. \'Oscar Pozo and Dr. Jon Romero for useful discussions.

\end{acknowledgments}

\bibliography{note}

\providecommand{\noopsort}[1]{}\providecommand{\singleletter}[1]{#1}%
\begin{thebibliography}{38}%
\makeatletter
\providecommand \@ifxundefined [1]{%
 \@ifx{#1\undefined}
}%
\providecommand \@ifnum [1]{%
 \ifnum #1\expandafter \@firstoftwo
 \else \expandafter \@secondoftwo
 \fi
}%
\providecommand \@ifx [1]{%
 \ifx #1\expandafter \@firstoftwo
 \else \expandafter \@secondoftwo
 \fi
}%
\providecommand \natexlab [1]{#1}%
\providecommand \enquote  [1]{``#1''}%
\providecommand \bibnamefont  [1]{#1}%
\providecommand \bibfnamefont [1]{#1}%
\providecommand \citenamefont [1]{#1}%
\providecommand \href@noop [0]{\@secondoftwo}%
\providecommand \href [0]{\begingroup \@sanitize@url \@href}%
\providecommand \@href[1]{\@@startlink{#1}\@@href}%
\providecommand \@@href[1]{\endgroup#1\@@endlink}%
\providecommand \@sanitize@url [0]{\catcode `\\12\catcode `\$12\catcode `\&12\catcode `\#12\catcode `\^12\catcode `\_12\catcode `\%12\relax}%
\providecommand \@@startlink[1]{}%
\providecommand \@@endlink[0]{}%
\providecommand \url  [0]{\begingroup\@sanitize@url \@url }%
\providecommand \@url [1]{\endgroup\@href {#1}{\urlprefix }}%
\providecommand \urlprefix  [0]{URL }%
\providecommand \Eprint [0]{\href }%
\providecommand \doibase [0]{http://dx.doi.org/}%
\providecommand \selectlanguage [0]{\@gobble}%
\providecommand \bibinfo  [0]{\@secondoftwo}%
\providecommand \bibfield  [0]{\@secondoftwo}%
\providecommand \translation [1]{[#1]}%
\providecommand \BibitemOpen [0]{}%
\providecommand \bibitemStop [0]{}%
\providecommand \bibitemNoStop [0]{.\EOS\space}%
\providecommand \EOS [0]{\spacefactor3000\relax}%
\providecommand \BibitemShut  [1]{\csname bibitem#1\endcsname}%
\let\auto@bib@innerbib\@empty
\bibitem [{\citenamefont {Fumagalli}\ \emph {et~al.}(2018)\citenamefont {Fumagalli}, \citenamefont {Esfandiar}, \citenamefont {Fabregas}, \citenamefont {Hu}, \citenamefont {Ares}, \citenamefont {Janardanan}, \citenamefont {Yang}, \citenamefont {Radha}, \citenamefont {Taniguchi}, \citenamefont {Watanabe} \emph {et~al.}}]{fumagalli2018}%
  \BibitemOpen
  \bibfield  {author} {\bibinfo {author} {\bibfnamefont {L.}~\bibnamefont {Fumagalli}}, \bibinfo {author} {\bibfnamefont {A.}~\bibnamefont {Esfandiar}}, \bibinfo {author} {\bibfnamefont {R.}~\bibnamefont {Fabregas}}, \bibinfo {author} {\bibfnamefont {S.}~\bibnamefont {Hu}}, \bibinfo {author} {\bibfnamefont {P.}~\bibnamefont {Ares}}, \bibinfo {author} {\bibfnamefont {A.}~\bibnamefont {Janardanan}}, \bibinfo {author} {\bibfnamefont {Q.}~\bibnamefont {Yang}}, \bibinfo {author} {\bibfnamefont {B.}~\bibnamefont {Radha}}, \bibinfo {author} {\bibfnamefont {T.}~\bibnamefont {Taniguchi}}, \bibinfo {author} {\bibfnamefont {K.}~\bibnamefont {Watanabe}},  \emph {et~al.},\ }\href {\doibase 10.1126/science.aat4191} {\bibfield  {journal} {\bibinfo  {journal} {Science}\ }\textbf {\bibinfo {volume} {360}},\ \bibinfo {pages} {1339} (\bibinfo {year} {2018})}\BibitemShut {NoStop}%
\bibitem [{\citenamefont {Hill}(1963)}]{Hill1963}%
  \BibitemOpen
  \bibfield  {author} {\bibinfo {author} {\bibfnamefont {N.~E.}\ \bibnamefont {Hill}},\ }\href {\doibase 10.1039/TF9635900344} {\bibfield  {journal} {\bibinfo  {journal} {Trans. Faraday Soc.}\ }\textbf {\bibinfo {volume} {59}},\ \bibinfo {pages} {344} (\bibinfo {year} {1963})}\BibitemShut {NoStop}%
\bibitem [{\citenamefont {Zhang}\ \emph {et~al.}(2013)\citenamefont {Zhang}, \citenamefont {Gygi},\ and\ \citenamefont {Galli}}]{Galli2013}%
  \BibitemOpen
  \bibfield  {author} {\bibinfo {author} {\bibfnamefont {C.}~\bibnamefont {Zhang}}, \bibinfo {author} {\bibfnamefont {F.}~\bibnamefont {Gygi}}, \ and\ \bibinfo {author} {\bibfnamefont {G.}~\bibnamefont {Galli}},\ }\href {\doibase 10.1021/jz401108n} {\bibfield  {journal} {\bibinfo  {journal} {J. Phys. Chem. Lett.}\ }\textbf {\bibinfo {volume} {4}},\ \bibinfo {pages} {2477} (\bibinfo {year} {2013})}\BibitemShut {NoStop}%
\bibitem [{\citenamefont {Zubeltzu}\ \emph {et~al.}(2016)\citenamefont {Zubeltzu}, \citenamefont {Corsetti}, \citenamefont {Fern{\'a}ndez-Serra},\ and\ \citenamefont {Artacho}}]{Zubeltzu2016}%
  \BibitemOpen
  \bibfield  {author} {\bibinfo {author} {\bibfnamefont {J.}~\bibnamefont {Zubeltzu}}, \bibinfo {author} {\bibfnamefont {F.}~\bibnamefont {Corsetti}}, \bibinfo {author} {\bibfnamefont {M.}~\bibnamefont {Fern{\'a}ndez-Serra}}, \ and\ \bibinfo {author} {\bibfnamefont {E.}~\bibnamefont {Artacho}},\ }\href {\doibase 10.1103/PhysRevE.93.062137} {\bibfield  {journal} {\bibinfo  {journal} {Phys. Rev. E}\ }\textbf {\bibinfo {volume} {93}},\ \bibinfo {pages} {062137} (\bibinfo {year} {2016})}\BibitemShut {NoStop}%
\bibitem [{\citenamefont {De~Luca}\ \emph {et~al.}(2016)\citenamefont {De~Luca}, \citenamefont {Kannam}, \citenamefont {Todd}, \citenamefont {Frascoli}, \citenamefont {Hansen},\ and\ \citenamefont {Daivis}}]{DeLuca2016}%
  \BibitemOpen
  \bibfield  {author} {\bibinfo {author} {\bibfnamefont {S.}~\bibnamefont {De~Luca}}, \bibinfo {author} {\bibfnamefont {S.~K.}\ \bibnamefont {Kannam}}, \bibinfo {author} {\bibfnamefont {B.}~\bibnamefont {Todd}}, \bibinfo {author} {\bibfnamefont {F.}~\bibnamefont {Frascoli}}, \bibinfo {author} {\bibfnamefont {J.~S.}\ \bibnamefont {Hansen}}, \ and\ \bibinfo {author} {\bibfnamefont {P.~J.}\ \bibnamefont {Daivis}},\ }\href {\doibase 10.1021/acs.langmuir.6b00791} {\bibfield  {journal} {\bibinfo  {journal} {Langmuir}\ }\textbf {\bibinfo {volume} {32}},\ \bibinfo {pages} {4765} (\bibinfo {year} {2016})}\BibitemShut {NoStop}%
\bibitem [{\citenamefont {Schlaich}\ \emph {et~al.}(2016)\citenamefont {Schlaich}, \citenamefont {Knapp},\ and\ \citenamefont {Netz}}]{Schlaich2016}%
  \BibitemOpen
  \bibfield  {author} {\bibinfo {author} {\bibfnamefont {A.}~\bibnamefont {Schlaich}}, \bibinfo {author} {\bibfnamefont {E.~W.}\ \bibnamefont {Knapp}}, \ and\ \bibinfo {author} {\bibfnamefont {R.~R.}\ \bibnamefont {Netz}},\ }\href {\doibase 10.1103/PhysRevLett.117.048001} {\bibfield  {journal} {\bibinfo  {journal} {Phys. Rev. Lett.}\ }\textbf {\bibinfo {volume} {117}},\ \bibinfo {pages} {048001} (\bibinfo {year} {2016})}\BibitemShut {NoStop}%
\bibitem [{\citenamefont {Zhang}(2018)}]{Zhang2018}%
  \BibitemOpen
  \bibfield  {author} {\bibinfo {author} {\bibfnamefont {C.}~\bibnamefont {Zhang}},\ }\href {\doibase 10.1063/1.5025150} {\bibfield  {journal} {\bibinfo  {journal} {J. Chem. Phys.}\ }\textbf {\bibinfo {volume} {148}} (\bibinfo {year} {2018}),\ 10.1063/1.5025150}\BibitemShut {NoStop}%
\bibitem [{\citenamefont {Monet}\ \emph {et~al.}(2021)\citenamefont {Monet}, \citenamefont {Bresme}, \citenamefont {Kornyshev},\ and\ \citenamefont {Berthoumieux}}]{Bresme2021}%
  \BibitemOpen
  \bibfield  {author} {\bibinfo {author} {\bibfnamefont {G.}~\bibnamefont {Monet}}, \bibinfo {author} {\bibfnamefont {F.}~\bibnamefont {Bresme}}, \bibinfo {author} {\bibfnamefont {A.}~\bibnamefont {Kornyshev}}, \ and\ \bibinfo {author} {\bibfnamefont {H.}~\bibnamefont {Berthoumieux}},\ }\href {\doibase 10.1103/PhysRevLett.126.216001} {\bibfield  {journal} {\bibinfo  {journal} {Phys. Rev. Lett.}\ }\textbf {\bibinfo {volume} {126}},\ \bibinfo {pages} {216001} (\bibinfo {year} {2021})}\BibitemShut {NoStop}%
\bibitem [{\citenamefont {Dufils}\ \emph {et~al.}(2024)\citenamefont {Dufils}, \citenamefont {Schran}, \citenamefont {Chen}, \citenamefont {Geim}, \citenamefont {Fumagalli},\ and\ \citenamefont {Michaelides}}]{Michaelides2024}%
  \BibitemOpen
  \bibfield  {author} {\bibinfo {author} {\bibfnamefont {T.}~\bibnamefont {Dufils}}, \bibinfo {author} {\bibfnamefont {C.}~\bibnamefont {Schran}}, \bibinfo {author} {\bibfnamefont {J.}~\bibnamefont {Chen}}, \bibinfo {author} {\bibfnamefont {A.~K.}\ \bibnamefont {Geim}}, \bibinfo {author} {\bibfnamefont {L.}~\bibnamefont {Fumagalli}}, \ and\ \bibinfo {author} {\bibfnamefont {A.}~\bibnamefont {Michaelides}},\ }\href {\doibase 10.1039/D3SC04740G} {\bibfield  {journal} {\bibinfo  {journal} {Chem. Sci.}\ }\textbf {\bibinfo {volume} {15}},\ \bibinfo {pages} {516} (\bibinfo {year} {2024})}\BibitemShut {NoStop}%
\bibitem [{\citenamefont {Dei{\ss}enbeck}\ and\ \citenamefont {Wippermann}(2023)}]{Deissenbeck2023}%
  \BibitemOpen
  \bibfield  {author} {\bibinfo {author} {\bibfnamefont {F.}~\bibnamefont {Dei{\ss}enbeck}}\ and\ \bibinfo {author} {\bibfnamefont {S.}~\bibnamefont {Wippermann}},\ }\href {\doibase 10.1021/acs.jctc.2c00959} {\bibfield  {journal} {\bibinfo  {journal} {J. Chem. Theo. Comp.}\ }\textbf {\bibinfo {volume} {19}},\ \bibinfo {pages} {1035} (\bibinfo {year} {2023})}\BibitemShut {NoStop}%
\bibitem [{\citenamefont {Olivieri}\ \emph {et~al.}(2021)\citenamefont {Olivieri}, \citenamefont {Hynes},\ and\ \citenamefont {Laage}}]{olivieri-JPCL-2021}%
  \BibitemOpen
  \bibfield  {author} {\bibinfo {author} {\bibfnamefont {J.-F.}\ \bibnamefont {Olivieri}}, \bibinfo {author} {\bibfnamefont {J.~T.}\ \bibnamefont {Hynes}}, \ and\ \bibinfo {author} {\bibfnamefont {D.}~\bibnamefont {Laage}},\ }\href {\doibase 10.1021/acs.jpclett.1c00447} {\bibfield  {journal} {\bibinfo  {journal} {J. Phys. Chem. Lett.}\ }\textbf {\bibinfo {volume} {12}},\ \bibinfo {pages} {4319} (\bibinfo {year} {2021})}\BibitemShut {NoStop}%
\bibitem [{\citenamefont {Motevaselian}\ and\ \citenamefont {Aluru}(2020)}]{Motevaselian-ACS-Nano-2020}%
  \BibitemOpen
  \bibfield  {author} {\bibinfo {author} {\bibfnamefont {M.~H.}\ \bibnamefont {Motevaselian}}\ and\ \bibinfo {author} {\bibfnamefont {N.~R.}\ \bibnamefont {Aluru}},\ }\href {\doibase 10.1021/acsnano.0c03173} {\bibfield  {journal} {\bibinfo  {journal} {ACS Nano}\ }\textbf {\bibinfo {volume} {14}},\ \bibinfo {pages} {12761} (\bibinfo {year} {2020})}\BibitemShut {NoStop}%
\bibitem [{\citenamefont {Ballenegger}\ and\ \citenamefont {Hansen}(2005)}]{Ballenegger2005}%
  \BibitemOpen
  \bibfield  {author} {\bibinfo {author} {\bibfnamefont {V.}~\bibnamefont {Ballenegger}}\ and\ \bibinfo {author} {\bibfnamefont {J.-P.}\ \bibnamefont {Hansen}},\ }\href {\doibase 10.1063/1.1845431} {\bibfield  {journal} {\bibinfo  {journal} {J. Chem. Phys.}\ }\textbf {\bibinfo {volume} {122}},\ \bibinfo {pages} {114711} (\bibinfo {year} {2005})}\BibitemShut {NoStop}%
\bibitem [{\citenamefont {Bonthuis}\ \emph {et~al.}(2011)\citenamefont {Bonthuis}, \citenamefont {Gekle},\ and\ \citenamefont {Netz}}]{Bonthuis2011}%
  \BibitemOpen
  \bibfield  {author} {\bibinfo {author} {\bibfnamefont {D.~J.}\ \bibnamefont {Bonthuis}}, \bibinfo {author} {\bibfnamefont {S.}~\bibnamefont {Gekle}}, \ and\ \bibinfo {author} {\bibfnamefont {R.~R.}\ \bibnamefont {Netz}},\ }\href {\doibase 10.1103/PhysRevLett.107.166102} {\bibfield  {journal} {\bibinfo  {journal} {Phys. Rev. Lett.}\ }\textbf {\bibinfo {volume} {107}},\ \bibinfo {pages} {166102} (\bibinfo {year} {2011})}\BibitemShut {NoStop}%
\bibitem [{\citenamefont {Boya}\ \emph {et~al.}(2024)\citenamefont {Boya}, \citenamefont {Keerthi},\ and\ \citenamefont {Parambath}}]{boya-phys-today-2024}%
  \BibitemOpen
  \bibfield  {author} {\bibinfo {author} {\bibfnamefont {R.}~\bibnamefont {Boya}}, \bibinfo {author} {\bibfnamefont {A.}~\bibnamefont {Keerthi}}, \ and\ \bibinfo {author} {\bibfnamefont {M.~S.}\ \bibnamefont {Parambath}},\ }\href {\doibase 10.1063/pt.frik.vxpk} {\bibfield  {journal} {\bibinfo  {journal} {Physics Today}\ }\textbf {\bibinfo {volume} {77}},\ \bibinfo {pages} {26} (\bibinfo {year} {2024})},\ \Eprint {http://arxiv.org/abs/https://pubs.aip.org/physicstoday/article-pdf/77/8/26/20082493/26\_1\_pt.frik.vxpk.pdf} {https://pubs.aip.org/physicstoday/article-pdf/77/8/26/20082493/26\_1\_pt.frik.vxpk.pdf} \BibitemShut {NoStop}%
\bibitem [{\citenamefont {Tian}\ \emph {et~al.}(2020)\citenamefont {Tian}, \citenamefont {Scullion}, \citenamefont {Hughes}, \citenamefont {Li}, \citenamefont {Shih}, \citenamefont {Coleman}, \citenamefont {Chhowalla},\ and\ \citenamefont {Santos}}]{Tian2020}%
  \BibitemOpen
  \bibfield  {author} {\bibinfo {author} {\bibfnamefont {T.}~\bibnamefont {Tian}}, \bibinfo {author} {\bibfnamefont {D.}~\bibnamefont {Scullion}}, \bibinfo {author} {\bibfnamefont {D.}~\bibnamefont {Hughes}}, \bibinfo {author} {\bibfnamefont {L.~H.}\ \bibnamefont {Li}}, \bibinfo {author} {\bibfnamefont {C.-J.}\ \bibnamefont {Shih}}, \bibinfo {author} {\bibfnamefont {J.}~\bibnamefont {Coleman}}, \bibinfo {author} {\bibfnamefont {M.}~\bibnamefont {Chhowalla}}, \ and\ \bibinfo {author} {\bibfnamefont {E.~J.}\ \bibnamefont {Santos}},\ }\href {\doibase 10.1021/acs.nanolett.9b02982} {\bibfield  {journal} {\bibinfo  {journal} {Nano Lett.}\ }\textbf {\bibinfo {volume} {20}},\ \bibinfo {pages} {841} (\bibinfo {year} {2020})}\BibitemShut {NoStop}%
\bibitem [{\citenamefont {Colombo}\ \emph {et~al.}(1991)\citenamefont {Colombo}, \citenamefont {Resta},\ and\ \citenamefont {Baroni}}]{Colombo1991}%
  \BibitemOpen
  \bibfield  {author} {\bibinfo {author} {\bibfnamefont {L.}~\bibnamefont {Colombo}}, \bibinfo {author} {\bibfnamefont {R.}~\bibnamefont {Resta}}, \ and\ \bibinfo {author} {\bibfnamefont {S.}~\bibnamefont {Baroni}},\ }\href {\doibase 10.1103/PhysRevB.44.5572} {\bibfield  {journal} {\bibinfo  {journal} {Phys. Rev. B}\ }\textbf {\bibinfo {volume} {44}},\ \bibinfo {pages} {5572} (\bibinfo {year} {1991})}\BibitemShut {NoStop}%
\bibitem [{\citenamefont {Bresme}\ and\ \citenamefont {Artacho}(2010)}]{Bresme2010}%
  \BibitemOpen
  \bibfield  {author} {\bibinfo {author} {\bibfnamefont {F.}~\bibnamefont {Bresme}}\ and\ \bibinfo {author} {\bibfnamefont {E.}~\bibnamefont {Artacho}},\ }\href {\doibase 10.1039/C0JM01572E} {\bibfield  {journal} {\bibinfo  {journal} {J. Mater. Chem.}\ }\textbf {\bibinfo {volume} {20}},\ \bibinfo {pages} {10351} (\bibinfo {year} {2010})}\BibitemShut {NoStop}%
\bibitem [{\citenamefont {Zhang}\ and\ \citenamefont {Sprik}(2016)}]{Sprik2016}%
  \BibitemOpen
  \bibfield  {author} {\bibinfo {author} {\bibfnamefont {C.}~\bibnamefont {Zhang}}\ and\ \bibinfo {author} {\bibfnamefont {M.}~\bibnamefont {Sprik}},\ }\href {\doibase 10.1103/PhysRevB.93.144201} {\bibfield  {journal} {\bibinfo  {journal} {Phys. Rev. B}\ }\textbf {\bibinfo {volume} {93}},\ \bibinfo {pages} {144201} (\bibinfo {year} {2016})}\BibitemShut {NoStop}%
\bibitem [{Note1()}]{Note1}%
  \BibitemOpen
  \bibinfo {note} {Both $\alpha _{\protect \mathrm {2D}}$ and $\alpha _{\protect \mathrm {3D}}$ are rank-two tensors but we are considering one element of the diagonal of the former ($\alpha _{\perp }$), and the latter is a number times the identity tensor for 3D bulk water.}\BibitemShut {Stop}%
\bibitem [{\citenamefont {Kumar}\ \emph {et~al.}(2005)\citenamefont {Kumar}, \citenamefont {Buldyrev}, \citenamefont {Starr}, \citenamefont {Giovambattista},\ and\ \citenamefont {Stanley}}]{Stanley}%
  \BibitemOpen
  \bibfield  {author} {\bibinfo {author} {\bibfnamefont {P.}~\bibnamefont {Kumar}}, \bibinfo {author} {\bibfnamefont {S.~V.}\ \bibnamefont {Buldyrev}}, \bibinfo {author} {\bibfnamefont {F.~W.}\ \bibnamefont {Starr}}, \bibinfo {author} {\bibfnamefont {N.}~\bibnamefont {Giovambattista}}, \ and\ \bibinfo {author} {\bibfnamefont {H.~E.}\ \bibnamefont {Stanley}},\ }\href {\doibase 10.1103/PhysRevE.72.051503} {\bibfield  {journal} {\bibinfo  {journal} {Physical Review E}\ }\textbf {\bibinfo {volume} {72}},\ \bibinfo {pages} {051503} (\bibinfo {year} {2005})}\BibitemShut {NoStop}%
\bibitem [{\citenamefont {Zangi}(2004)}]{zangi2004water}%
  \BibitemOpen
  \bibfield  {author} {\bibinfo {author} {\bibfnamefont {R.}~\bibnamefont {Zangi}},\ }\href {\doibase 10.1088/0953-8984/16/45/005} {\bibfield  {journal} {\bibinfo  {journal} {Journal of Physics: Condensed Matter}\ }\textbf {\bibinfo {volume} {16}},\ \bibinfo {pages} {S5371} (\bibinfo {year} {2004})}\BibitemShut {NoStop}%
\bibitem [{\citenamefont {Calero}\ and\ \citenamefont {Franzese}(2020)}]{calero2020water}%
  \BibitemOpen
  \bibfield  {author} {\bibinfo {author} {\bibfnamefont {C.}~\bibnamefont {Calero}}\ and\ \bibinfo {author} {\bibfnamefont {G.}~\bibnamefont {Franzese}},\ }\href {\doibase https://doi.org/10.1016/j.molliq.2020.114027} {\bibfield  {journal} {\bibinfo  {journal} {Journal of Molecular Liquids}\ }\textbf {\bibinfo {volume} {317}},\ \bibinfo {pages} {114027} (\bibinfo {year} {2020})}\BibitemShut {NoStop}%
\bibitem [{\citenamefont {Leoni}\ \emph {et~al.}(2021)\citenamefont {Leoni}, \citenamefont {Calero},\ and\ \citenamefont {Franzese}}]{leoni2021nanoconfined}%
  \BibitemOpen
  \bibfield  {author} {\bibinfo {author} {\bibfnamefont {F.}~\bibnamefont {Leoni}}, \bibinfo {author} {\bibfnamefont {C.}~\bibnamefont {Calero}}, \ and\ \bibinfo {author} {\bibfnamefont {G.}~\bibnamefont {Franzese}},\ }\href {\doibase https://doi.org/10.1021/acsnano.1c07381} {\bibfield  {journal} {\bibinfo  {journal} {Acs Nano}\ }\textbf {\bibinfo {volume} {15}},\ \bibinfo {pages} {19864} (\bibinfo {year} {2021})}\BibitemShut {NoStop}%
\bibitem [{\citenamefont {Han}\ \emph {et~al.}(2010)\citenamefont {Han}, \citenamefont {Choi}, \citenamefont {Kumar},\ and\ \citenamefont {Stanley}}]{han2010phase}%
  \BibitemOpen
  \bibfield  {author} {\bibinfo {author} {\bibfnamefont {S.}~\bibnamefont {Han}}, \bibinfo {author} {\bibfnamefont {M.}~\bibnamefont {Choi}}, \bibinfo {author} {\bibfnamefont {P.}~\bibnamefont {Kumar}}, \ and\ \bibinfo {author} {\bibfnamefont {H.~E.}\ \bibnamefont {Stanley}},\ }\href {\doibase https://doi.org/10.1038/nphys1708} {\bibfield  {journal} {\bibinfo  {journal} {Nature Physics}\ }\textbf {\bibinfo {volume} {6}},\ \bibinfo {pages} {685} (\bibinfo {year} {2010})}\BibitemShut {NoStop}%
\bibitem [{\citenamefont {Abascal}\ and\ \citenamefont {Vega}(2005)}]{TIP4P}%
  \BibitemOpen
  \bibfield  {author} {\bibinfo {author} {\bibfnamefont {J.~L.}\ \bibnamefont {Abascal}}\ and\ \bibinfo {author} {\bibfnamefont {C.}~\bibnamefont {Vega}},\ }\href {\doibase 10.1063/1.2121687} {\bibfield  {journal} {\bibinfo  {journal} {J. Chem. Phys.}\ }\textbf {\bibinfo {volume} {123}} (\bibinfo {year} {2005}),\ 10.1063/1.2121687}\BibitemShut {NoStop}%
\bibitem [{Note2()}]{Note2}%
  \BibitemOpen
  \bibinfo {note} {The 2D molecular density range of 0.177 \r A$^{-2}$ - 0.200 \r A$^{-2}$ corresponds to a 3D mass density range of 1.04 g/cm$^3$ - 1.17 g/cm$^3$ in Ref.\cite {Zubeltzu2016}}\BibitemShut {NoStop}%
\bibitem [{\citenamefont {Thompson}\ \emph {et~al.}(2022)\citenamefont {Thompson}, \citenamefont {Aktulga}, \citenamefont {Berger}, \citenamefont {Bolintineanu}, \citenamefont {Brown}, \citenamefont {Crozier}, \citenamefont {in~'t Veld}, \citenamefont {Kohlmeyer}, \citenamefont {Moore}, \citenamefont {Nguyen}, \citenamefont {Shan}, \citenamefont {Stevens}, \citenamefont {Tranchida}, \citenamefont {Trott},\ and\ \citenamefont {Plimpton}}]{LAMMPS}%
  \BibitemOpen
  \bibfield  {author} {\bibinfo {author} {\bibfnamefont {A.~P.}\ \bibnamefont {Thompson}}, \bibinfo {author} {\bibfnamefont {H.~M.}\ \bibnamefont {Aktulga}}, \bibinfo {author} {\bibfnamefont {R.}~\bibnamefont {Berger}}, \bibinfo {author} {\bibfnamefont {D.~S.}\ \bibnamefont {Bolintineanu}}, \bibinfo {author} {\bibfnamefont {W.~M.}\ \bibnamefont {Brown}}, \bibinfo {author} {\bibfnamefont {P.~S.}\ \bibnamefont {Crozier}}, \bibinfo {author} {\bibfnamefont {P.~J.}\ \bibnamefont {in~'t Veld}}, \bibinfo {author} {\bibfnamefont {A.}~\bibnamefont {Kohlmeyer}}, \bibinfo {author} {\bibfnamefont {S.~G.}\ \bibnamefont {Moore}}, \bibinfo {author} {\bibfnamefont {T.~D.}\ \bibnamefont {Nguyen}}, \bibinfo {author} {\bibfnamefont {R.}~\bibnamefont {Shan}}, \bibinfo {author} {\bibfnamefont {M.~J.}\ \bibnamefont {Stevens}}, \bibinfo {author} {\bibfnamefont {J.}~\bibnamefont {Tranchida}}, \bibinfo {author} {\bibfnamefont {C.}~\bibnamefont {Trott}}, \ and\ \bibinfo {author} {\bibfnamefont {S.~J.}\ \bibnamefont {Plimpton}},\
  }\href {\doibase 10.1016/j.cpc.2021.108171} {\bibfield  {journal} {\bibinfo  {journal} {Comp. Phys. Comm.}\ }\textbf {\bibinfo {volume} {271}},\ \bibinfo {pages} {108171} (\bibinfo {year} {2022})}\BibitemShut {NoStop}%
\bibitem [{\citenamefont {Soler}\ \emph {et~al.}(2002)\citenamefont {Soler}, \citenamefont {Artacho}, \citenamefont {Gale}, \citenamefont {Garc\'{\i}a}, \citenamefont {Junquera}, \citenamefont {Ordej\'on},\ and\ \citenamefont {S\'anchez-Portal}}]{Soler2002}%
  \BibitemOpen
  \bibfield  {author} {\bibinfo {author} {\bibfnamefont {J.~M.}\ \bibnamefont {Soler}}, \bibinfo {author} {\bibfnamefont {E.}~\bibnamefont {Artacho}}, \bibinfo {author} {\bibfnamefont {J.~D.}\ \bibnamefont {Gale}}, \bibinfo {author} {\bibfnamefont {A.}~\bibnamefont {Garc\'{\i}a}}, \bibinfo {author} {\bibfnamefont {J.}~\bibnamefont {Junquera}}, \bibinfo {author} {\bibfnamefont {P.}~\bibnamefont {Ordej\'on}}, \ and\ \bibinfo {author} {\bibfnamefont {D.}~\bibnamefont {S\'anchez-Portal}},\ }\href {\doibase 10.1088/0953-8984/14/11/302} {\bibfield  {journal} {\bibinfo  {journal} {J. Phys.: Condens. Matter}\ }\textbf {\bibinfo {volume} {14}},\ \bibinfo {pages} {2745} (\bibinfo {year} {2002})}\BibitemShut {NoStop}%
\bibitem [{\citenamefont {Perdew}\ \emph {et~al.}(1996)\citenamefont {Perdew}, \citenamefont {Burke},\ and\ \citenamefont {Ernzerhof}}]{PBE2}%
  \BibitemOpen
  \bibfield  {author} {\bibinfo {author} {\bibfnamefont {J.~P.}\ \bibnamefont {Perdew}}, \bibinfo {author} {\bibfnamefont {K.}~\bibnamefont {Burke}}, \ and\ \bibinfo {author} {\bibfnamefont {M.}~\bibnamefont {Ernzerhof}},\ }\href {\doibase 10.1103/PhysRevLett.77.3865} {\bibfield  {journal} {\bibinfo  {journal} {Phys. Rev. Lett.}\ }\textbf {\bibinfo {volume} {77}},\ \bibinfo {pages} {3865} (\bibinfo {year} {1996})}\BibitemShut {NoStop}%
\bibitem [{\citenamefont {Corsetti}\ \emph {et~al.}(2013)\citenamefont {Corsetti}, \citenamefont {Fern{\'a}ndez-Serra}, \citenamefont {Soler},\ and\ \citenamefont {Artacho}}]{corsetti2013b}%
  \BibitemOpen
  \bibfield  {author} {\bibinfo {author} {\bibfnamefont {F.}~\bibnamefont {Corsetti}}, \bibinfo {author} {\bibfnamefont {M.}~\bibnamefont {Fern{\'a}ndez-Serra}}, \bibinfo {author} {\bibfnamefont {J.~M.}\ \bibnamefont {Soler}}, \ and\ \bibinfo {author} {\bibfnamefont {E.}~\bibnamefont {Artacho}},\ }\href {\doibase 10.1088/0953-8984/25/43/435504} {\bibfield  {journal} {\bibinfo  {journal} {Journal of Physics: Condensed Matter}\ }\textbf {\bibinfo {volume} {25}},\ \bibinfo {pages} {435504} (\bibinfo {year} {2013})}\BibitemShut {NoStop}%
\bibitem [{\citenamefont {Ge}\ and\ \citenamefont {Lu}(2017)}]{molecule_pol}%
  \BibitemOpen
  \bibfield  {author} {\bibinfo {author} {\bibfnamefont {X.}~\bibnamefont {Ge}}\ and\ \bibinfo {author} {\bibfnamefont {D.}~\bibnamefont {Lu}},\ }\href {\doibase 10.1103/PhysRevB.96.075114} {\bibfield  {journal} {\bibinfo  {journal} {Phys. Rev. B}\ }\textbf {\bibinfo {volume} {96}},\ \bibinfo {pages} {075114} (\bibinfo {year} {2017})}\BibitemShut {NoStop}%
\bibitem [{\citenamefont {Haynes}(2016)}]{Handbook}%
  \BibitemOpen
  \bibfield  {author} {\bibinfo {author} {\bibfnamefont {W.~M.}\ \bibnamefont {Haynes}},\ }\href {\doibase 10.1201/9781315380476} {\emph {\bibinfo {title} {{CRC Handbook of Chemistry and Physics}}}}\ (\bibinfo  {publisher} {CRC press},\ \bibinfo {year} {2016})\BibitemShut {NoStop}%
\bibitem [{\citenamefont {Manino}\ \emph {et~al.}(2024)\citenamefont {Manino}, \citenamefont {Arvelos}, \citenamefont {Kaushik}, \citenamefont {Ordejon}, \citenamefont {Rocha}, \citenamefont {Pedroza},\ and\ \citenamefont {Fernández-Serra}}]{Marivi2}%
  \BibitemOpen
  \bibfield  {author} {\bibinfo {author} {\bibfnamefont {A.}~\bibnamefont {Manino}}, \bibinfo {author} {\bibfnamefont {G.~M.}\ \bibnamefont {Arvelos}}, \bibinfo {author} {\bibfnamefont {K.}~\bibnamefont {Kaushik}}, \bibinfo {author} {\bibfnamefont {P.}~\bibnamefont {Ordejon}}, \bibinfo {author} {\bibfnamefont {A.~R.}\ \bibnamefont {Rocha}}, \bibinfo {author} {\bibfnamefont {L.~S.}\ \bibnamefont {Pedroza}}, \ and\ \bibinfo {author} {\bibfnamefont {M.}~\bibnamefont {Fernández-Serra}},\ }\href@noop {} {\bibfield  {journal} {\bibinfo  {journal} {In prep}\ } (\bibinfo {year} {2024})}\BibitemShut {NoStop}%
\bibitem [{\citenamefont {Becker}\ \emph {et~al.}(2023)\citenamefont {Becker}, \citenamefont {Loche}, \citenamefont {Rezaei}, \citenamefont {Wolde-Kidan}, \citenamefont {Uematsu}, \citenamefont {Netz},\ and\ \citenamefont {Bonthuis}}]{becker2023multiscale}%
  \BibitemOpen
  \bibfield  {author} {\bibinfo {author} {\bibfnamefont {M.}~\bibnamefont {Becker}}, \bibinfo {author} {\bibfnamefont {P.}~\bibnamefont {Loche}}, \bibinfo {author} {\bibfnamefont {M.}~\bibnamefont {Rezaei}}, \bibinfo {author} {\bibfnamefont {A.}~\bibnamefont {Wolde-Kidan}}, \bibinfo {author} {\bibfnamefont {Y.}~\bibnamefont {Uematsu}}, \bibinfo {author} {\bibfnamefont {R.~R.}\ \bibnamefont {Netz}}, \ and\ \bibinfo {author} {\bibfnamefont {D.~J.}\ \bibnamefont {Bonthuis}},\ }\href {\doibase https://doi.org/10.1021/acs.chemrev.3c00307} {\bibfield  {journal} {\bibinfo  {journal} {Chemical Reviews}\ }\textbf {\bibinfo {volume} {124}},\ \bibinfo {pages} {1} (\bibinfo {year} {2023})}\BibitemShut {NoStop}%
\bibitem [{\citenamefont {Loche}\ \emph {et~al.}(2020)\citenamefont {Loche}, \citenamefont {Ayaz}, \citenamefont {Wolde-Kidan}, \citenamefont {Schlaich},\ and\ \citenamefont {Netz}}]{loche2020universal}%
  \BibitemOpen
  \bibfield  {author} {\bibinfo {author} {\bibfnamefont {P.}~\bibnamefont {Loche}}, \bibinfo {author} {\bibfnamefont {C.}~\bibnamefont {Ayaz}}, \bibinfo {author} {\bibfnamefont {A.}~\bibnamefont {Wolde-Kidan}}, \bibinfo {author} {\bibfnamefont {A.}~\bibnamefont {Schlaich}}, \ and\ \bibinfo {author} {\bibfnamefont {R.~R.}\ \bibnamefont {Netz}},\ }\href {\doibase https://doi.org/10.1021/acs.jpcb.0c01967} {\bibfield  {journal} {\bibinfo  {journal} {The Journal of Physical Chemistry B}\ }\textbf {\bibinfo {volume} {124}},\ \bibinfo {pages} {4365} (\bibinfo {year} {2020})}\BibitemShut {NoStop}%
\bibitem [{\citenamefont {Nir}\ \emph {et~al.}(1973)\citenamefont {Nir}, \citenamefont {Adams},\ and\ \citenamefont {Rein}}]{nir1973polarizability}%
  \BibitemOpen
  \bibfield  {author} {\bibinfo {author} {\bibfnamefont {S.}~\bibnamefont {Nir}}, \bibinfo {author} {\bibfnamefont {S.}~\bibnamefont {Adams}}, \ and\ \bibinfo {author} {\bibfnamefont {R.}~\bibnamefont {Rein}},\ }\href {\doibase https://doi.org/10.1063/1.1680478} {\bibfield  {journal} {\bibinfo  {journal} {The Journal of Chemical Physics}\ }\textbf {\bibinfo {volume} {59}},\ \bibinfo {pages} {3341} (\bibinfo {year} {1973})}\BibitemShut {NoStop}%
\bibitem [{\citenamefont {Wen}\ \emph {et~al.}(2023)\citenamefont {Wen}, \citenamefont {Ma}, \citenamefont {Mannino}, \citenamefont {Fernandez-Serra}, \citenamefont {Shen},\ and\ \citenamefont {Catalan}}]{Marivi1}%
  \BibitemOpen
  \bibfield  {author} {\bibinfo {author} {\bibfnamefont {X.}~\bibnamefont {Wen}}, \bibinfo {author} {\bibfnamefont {Q.}~\bibnamefont {Ma}}, \bibinfo {author} {\bibfnamefont {A.}~\bibnamefont {Mannino}}, \bibinfo {author} {\bibfnamefont {M.}~\bibnamefont {Fernandez-Serra}}, \bibinfo {author} {\bibfnamefont {S.}~\bibnamefont {Shen}}, \ and\ \bibinfo {author} {\bibfnamefont {G.}~\bibnamefont {Catalan}},\ }\href@noop {} {\bibfield  {journal} {\bibinfo  {journal} {arXiv preprint arXiv:2212.00323}\ } (\bibinfo {year} {2023})}\BibitemShut {NoStop}%
\end{thebibliography}%

\end{document}


\maketitle

\clearpage 
\section*{S1: Molecular polarisability of a water monomer}

\begin{figure*}[h]
\centering
\subfigure[]{\includegraphics[width=0.45\textwidth]{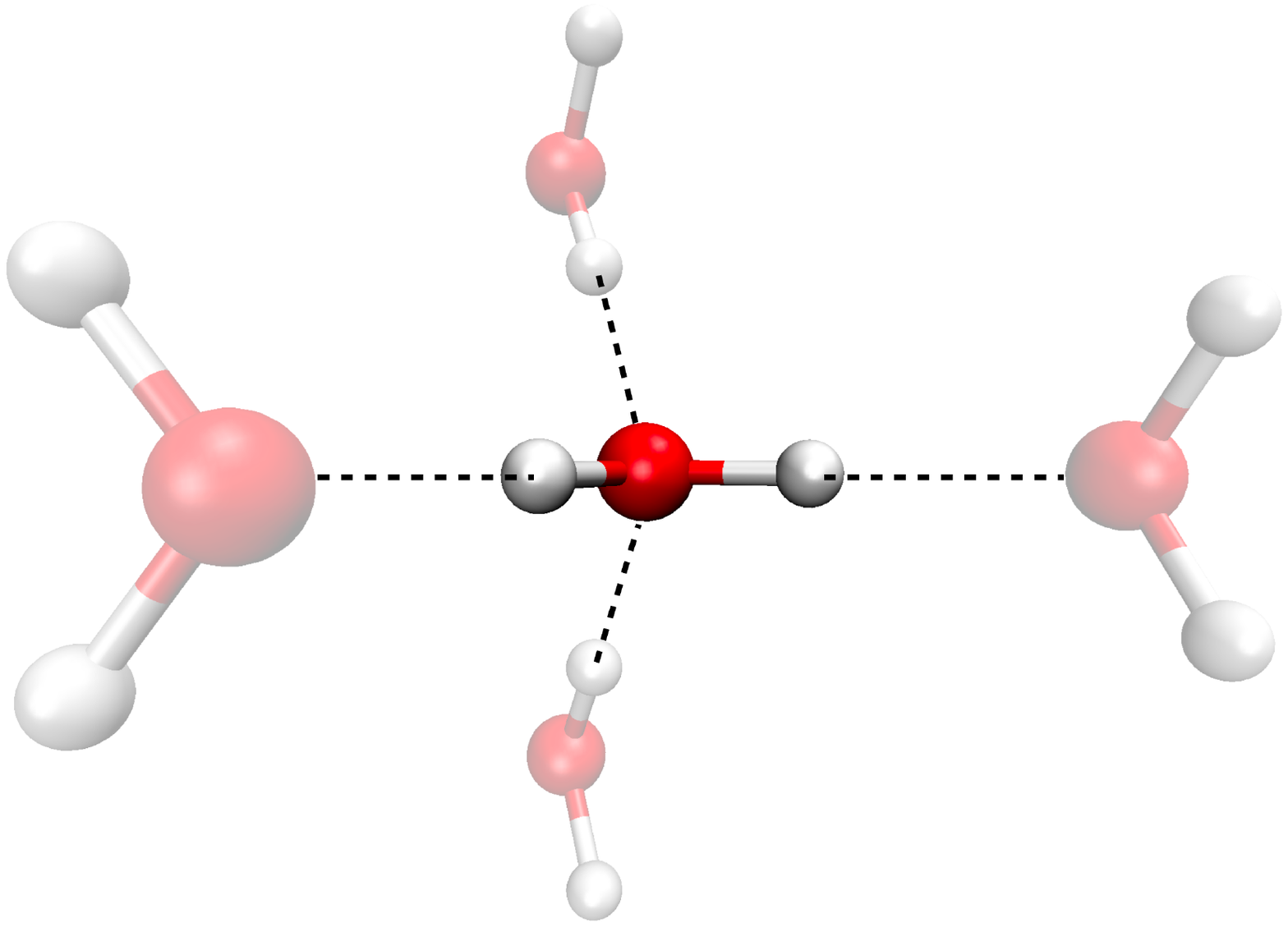}}
\hspace{0.05\textwidth}
\subfigure[]{\includegraphics[width=0.45\textwidth]{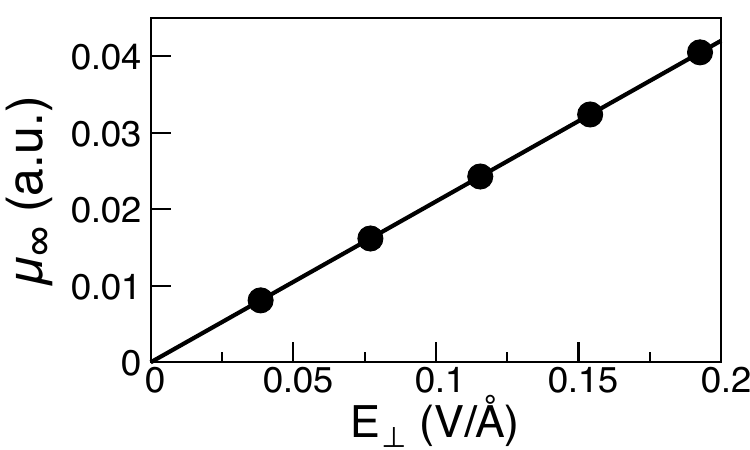}}
\caption{(a) Molecular positions obtained from the I$_{h}$ ice structure. The electric field is applied perpendicular to the plane defined by the atomic positions of the central water molecule. (b) Dipole moment of the central molecule as a function of the applied electric field. The solid line represents a molecular polarizability of 1.60 \AA$^3$ in Gaussian units.}
\label{fig:molecular_pol}
\end{figure*}

In order to assess the accuracy of our simulation setup, we calculate the electronic polarisability of a water monomer. Figure S1 (a) illustrates the atomic configuration used in these calculations. The system consists of five water molecules: a central molecule surrounded by four others. The latter are treated as ghost atoms, providing a larger basis set, while not contributing to the dipole moment of the system. The atomic configurations are taken from the I$_h$ ice structure. The external field is applied along the direction perpendicular to the plane defined by the atomic positions of the central molecule, ensuring that there is no contribution from the permanent dipole moment. Previous simulations employing the same exchange-correlation functional indicate that the anisotropy of the electronic polarisability of a water molecule is small compared to its mean value.

Figure \ref{fig:molecular_pol} (b) shows the dipole moment of the central molecule for different values of the applied external field. From the linear relationship between these two quantities, we obtain a polarisability of 1.60 \AA$^3$ in Gaussian units for a water monomer.

\clearpage

\section*{S2: Dipole moment orientation}

\begin{figure*}[h]
\centering
\includegraphics[height=8.0cm]{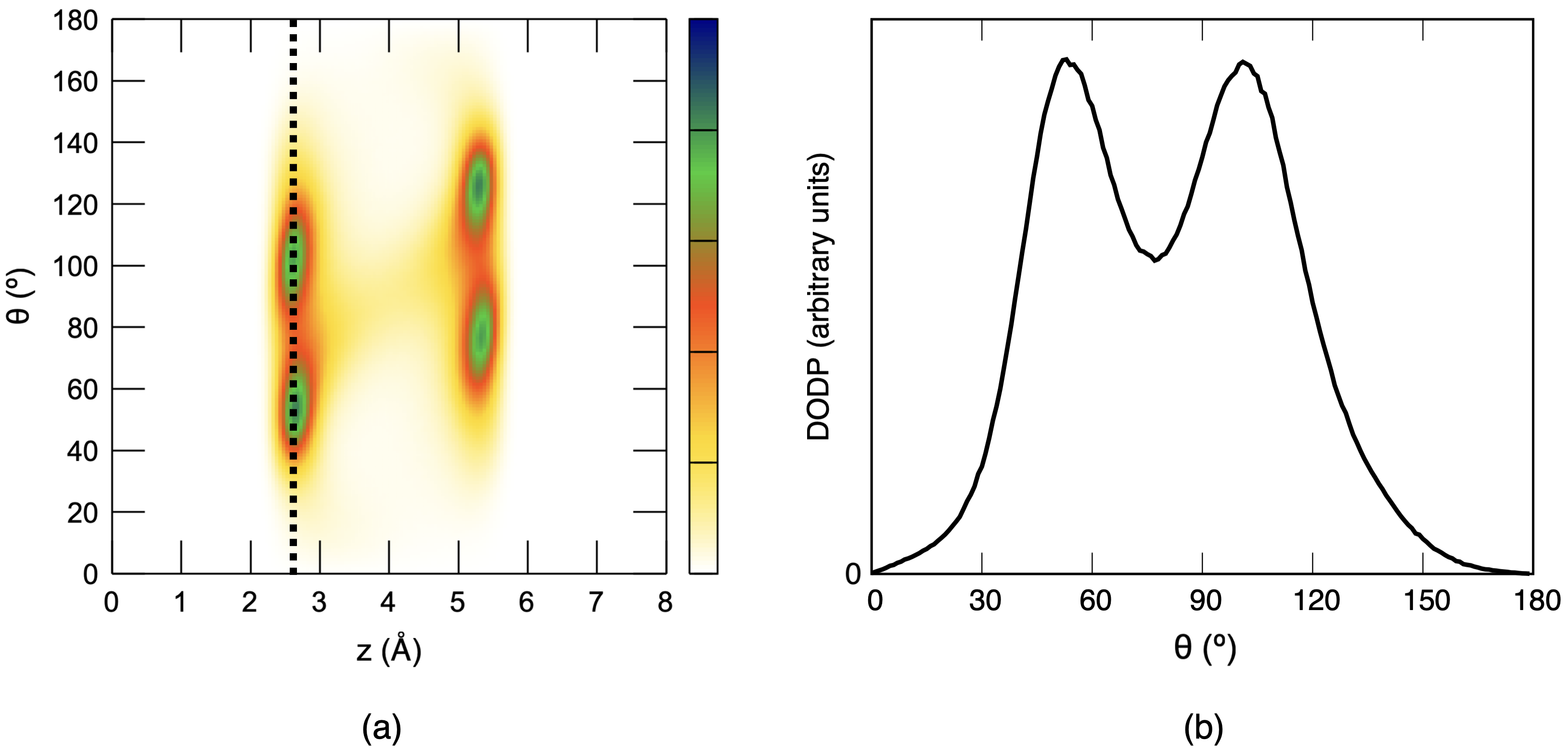}
\caption{(a) Histogram of dipole moment orientations of water molecules along the confining direction for the TIP4P/2005 model. (b) Dipole orientation distribution profile of water molecules at \(z = 2.6 \, \text{\AA}\) (indicated by the dashed vertical line in (a)).}
\label{fig:SI_ADF}
\end{figure*}

Figure \ref{fig:SI_ADF}(a) presents the histogram of dipole moment orientations of water molecules relative to the positive direction of the $z$-axis (the confining direction) for the TIP4P/2005 model in the absence of an external electric field. Figure \ref{fig:SI_ADF}(b) displays the dipole orientation distribution profile (DODP) for the same water molecules at $z$ = 2.6 \AA, corresponding to the dashed vertical line in Figure \ref{fig:SI_ADF}(a). 
\clearpage 
\bibliographystyle{unsrt}  
\bibliography{note}